\documentclass[aps, prl, superscriptaddress, preprintnumbers, twocolumn]{revtex4}
\usepackage{amsmath}	% AMS Math Package
\usepackage{amsthm}	% Theorem Formatting
\usepackage{amssymb}	% Math symbols such as \mathbb
\usepackage{datetime}
\usepackage{graphicx}
\usepackage{color}
\usepackage{verbatim}

\usepackage{hyperref}

\definecolor{grey}{rgb}{0.4,0.4,0.4}
\definecolor{dullmagenta}{rgb}{0.4,0,0.4}
\definecolor{darkblue}{rgb}{0,0,0.4}
\definecolor{midblue}{rgb}{0,0,0.5}
\definecolor{midred}{rgb}{0.5,0,0}
\definecolor{orange}{rgb}{1,0.5,0}
\definecolor{lightbrown}{rgb}{0.75,0.5,0.25}
\definecolor{tan}{cmyk}{0.14,0.42,0.56,0}
\definecolor{djunglegreen}{cmyk}{0.99,0,0.52,0}
\definecolor{lightgreen}{rgb}{0,1,0}
\definecolor{olivegreen}{cmyk}{0.64,0,0.95,0.40}
\definecolor{midgreen}{rgb}{0.0,0.675,0.0}
\definecolor{darkgreen}{rgb}{0,0.5,0}

\newcommand{\be}{\begin{equation}}
\newcommand{\ee}{\end{equation}}

\newcommand{\vs}{\vspace}

\renewcommand{\.}{\hspace{0.5mm}}

\renewcommand{\d}{\ensuremath{\mathrm{d}}}
\newcommand{\eg}{e.g.}
\newcommand{\ie}{i.e.}

\newcommand{\cf}{cf.} % CR: according to Oxford English, ``c.f.'' means carried forward

\let\baraccent=\= % rename builtin command \= to \baraccent
\renewcommand{\=}[1]{\stackrel{#1}{=}} % for putting numbers above =

\theoremstyle{definition}

\theoremstyle{remark}

\settimeformat{ampmtime}

\usepackage{float}

\begin{document}

\title{Primordial Black-Hole and Macroscopic Dark-Matter Constraints with LISA}

\author{Florian K{\"u}hnel}
\email{florian.kuhnel@fysik.su.se}
\affiliation{The Oskar Klein Centre for Cosmoparticle Physics,
	Department of Physics,
	Stockholm University,
	AlbaNova University Center,
	Roslagstullsbacken 21,
	SE--106\.91 Stockholm,
	Sweden}

\affiliation{Department of Physics,
	School of Engineering Sciences,
	KTH Royal Institute of Technology,
	AlbaNova University Center,
	Roslagstullsbacken 21,
	SE--106\.91 Stockholm,
	Sweden}

\author{Glenn D.~Starkman}
\email{glenn.starkman@case.edu}
\affiliation{CERCA/ISO,
	Department of Physics,
	Case Western Reserve University,
	10900 Euclid Avenue, 
	Cleveland, 
	OH 44106, 
	USA}

\author{Katherine Freese}
\email{ktfreese@umich.edu}
\affiliation{Department of Physics,
	University of Michigan,
	Ann Arbor,
	MI 48109,
	USA}

\affiliation{The Oskar Klein Centre for Cosmoparticle Physics,
	Department of Physics,
	Stockholm University,
	AlbaNova University Center,
	Roslagstullsbacken 21,
	SE--106\.91 Stockholm,
	Sweden}

\author{Andrew Matas}
\email{amatas@umn.edu}
\affiliation{University of Minnesota,
	Minneapolis,
	MN 55455,
	USA}

\date{\formatdate{\day}{\month}{\year}, \currenttime}

%%%%%%%%%%%%%%%%%%%%%%%%%%%%%%%%%%%%%%%%%%%%%%%%%%%%%%%%%%%%

\begin{abstract}
The detectability of gravitational waves originating from primordial black holes or other large macroscopic dark-matter candidates inspiraling into Sagittarius ${\rm A}^{\!*}$ is investigated. 
It is shown that LISA should be a formidable machine to detect the stochastic signal from the accumulated collection of such objects that have been gravitationally captured by Sgr\.${\rm A}^{\!*}$, provided they have masses above $\sim 10^{-20}\.M_{\odot}$.
This is especially so, if they are the main component of the dark matter, and the dark-matter density at Sagittarius ${\rm A}^{\!*}$ exceeds its value in the solar-neighborhood.
Notably, the window of mass that is unconstrained by various gravitational lensing arguments and observations, $2 \times 10^{20}\.{\rm g} \leq m_{\rm DM} \leq 4 \times 10^{24}\.{\rm g}$, will be accessible. 
Forecasts are derived for the event rates and signal strengths as a function of dark-matter mass, assuming that the dark-matter candidates are not tidally disrupted during the inspiral. 
This is certainly the case for primordial black holes, and it shown that it is also very likely the case for candidate objects of nuclear density.
\end{abstract}

\maketitle

%%%%%%%%%%%%%%%%%%%%%%%%%%%%%%%%%%%%%%%%%%%%%%%%%%%%%%%%%%%%

According to the current standard cosmological model, approximately $25\.$\% of the energy density of the Universe is in the form of so-called cold dark matter{\,---\,}non-relativistic masses forming, at least approximately, a perfect fluid of negligible pressure. 
The leading candidate has long been hypothetical heavy ($m \gtrsim 1\.$GeV) particles outside the Standard Model of particle physics, possessing very small scattering cross-sections on each other and on Standard-Model particles. 
These are known for short as WIMPs.
However, it has also long-been recognized that there are viable dark-matter candidates of much greater mass and cross-section, notably primordial black holes (PBH) \cite{1967SvA....10..602Z, Carr:1974nx} (see also Refs.~\cite{Jedamzik:1996mr, Niemeyer:1997mt, Jedamzik:2000ap, Musco:2004ak, Musco:2008hv, Capela:2012jz, Griest:2013aaa, Belotsky:2014kca, Young:2015kda, Frampton:2015xza, Bird:2016dcv, Kawasaki:2016pql, Carr:2016drx, Kashlinsky:2016sdv, Clesse:2016vqa, Green:2016xgy, Kuhnel:2017pwq, Akrami:2016vrq, Garcia-Bellido:2017fdg, Garcia-Bellido:2017qal, Carr:2017jsz}) 
and objects of nuclear density (\eg~Ref.~\cite{Witten:1984rs, Lynn:1989xb, Kusenko:1997si, Zhitnitsky:2002nr, Lynn:2010uh}), either of which could potentially be the result of Standard-Model physics in the early Universe.
For the purposes of this paper, we will refer to all such macroscopic dark-matter candidates, including PBHs, generically as macros.

Meanwhile, we have recently seen the dawn of gravitational-wave astronomy \cite{Abbott:2016blz}. 
Advanced LIGO, operating in approximately the $1$ - $1000\.$Hz band, is now regularly detecting the merger of black holes that are of tens of solar masses. 
It is expected that it will sometime soon also detect neutron-star mergers.

With the continued non-detection of WIMP dark matter, and the failure of long-predicted Beyond the Standard-Model physics to materialize at the Large Hadron Collider, the case for alternative, and especially Standard Model, candidates has grown stronger, and attracted increasing attention. 
There has been significant attention to the possibility that multi-solar-mass black holes, such as those detected by Advanced LIGO, could be the dark matter \cite{Kovetz:2016kpi}.

The anticipated next generation of gravitational-wave detectors, such as eLISA or LISA will operate in the $0.1$ - $100$\.mHz band, and is expected to be sensitive to supermassive black-hole binaries, Galactic white-dwarf binaries, and extreme mass-ratio inspirals (EMRI) \cite{Moore:2014lga}.
Clesse {\it et al.} \cite{Clesse:2016ajp} have suggested that if the dark matter is composed of multi-solar-mass black holes that LISA could detect their merger.

Lighter macros are also of definite interest.
There are well-known limits from microlensing of Milky Way and Magellanic Cloud stars \cite{Allsman:2000kg, Tisserand:2006zx, Carr:2009jm, PhysRevLett.111.181302} limiting the abundance of macros above approximately $4 \times 10^{24}\.{\rm g}$.
The failure to observe femtolensing of gamma-ray bursts \cite{1992ApJ...386L...5G} means that dark matter cannot be composed entirely of macros between approximately $2 \times 10^{17}\.{\rm g}$ and $2 \times 10^{20}\.{\rm g}$.
However, between about $2 \times 10^{20}\.{\rm g}$ and $4 \times 10^{24}\.{\rm g}$ there is an unconstrained window for anything of approximately ordinary matter density or greater \cite{Jacobs:2014yca}. 
Candidates of approximately nuclear or greater density are also unconstrained from $55\.{\rm g}$ to $2 \times 10^{17}\.{\rm g}$ \cite{Jacobs:2014yca}, although if, as expected, primordial black holes emit Hawking radiation, then they would have evaporated before now if their masses were below approximately $10^{15}\.{\rm g}$. 

If the dark matter is indeed composed of dense objects with $m \ll M_{\odot} \simeq 2 \times 10^{33}\.{\rm g}$, then their capture by a supermassive black hole SMBH is an extreme mass-ratio inspiral (EMRI) event, and the upcoming space-based gravity waves detectors are potential detectors of this broad class of dark matter. 
The most promising target is the Sagittarius ${\rm A}^{\!*}$ SMBH at the center of our Galaxy.
Even though the strain associated with the gravitational waves from an individual macro inspiraling into Sgr\.${\rm A}^{\!*}$ is likely to be below the sensitivity threshold of any near-term gravitational-wave detectors such as LISA, the inspiral of a captured macro will typically take longer than the age of the Universe. 
Sgr\.${\rm A}^{\!*}$ will therefore have accumulated a large cloud of macros that are all slowly spiralling inward, emitting gravitational waves as they go.
In this letter, we argue that the collective stochastic signal due to these ongoing inspirals is potentially detectable.

We now proceed to estimate the expected rate and signal strengths of gravitational-wave EMRI events from macro/PBH encounters with the Sagittarius ${\rm A}^{\!*}$ SMBH. 

Let $m$ be the mass of the inspiraling macro, $M \gg m$ the mass of the SMBH with which it merges, and $v$ their relative velocity. 
The minimum impact parameter is labelled $b_{\rm min}$, which is connected to the merger cross-section $\sigma_{\rm mer} \equiv \pi\.b_{\rm min}^{2}$. 
Ref.~\cite{Mouri:2002mc} has shown 
\begin{align}
	\sigma_{\rm mer}
		&=
				2\pi \left( \frac{85\.\pi}{6\.\sqrt{2\,}} \right)^{\!\!2 / 7}
				\frac{G^{2}\.( m + M )^{10 / 7} m^{2 / 7} M^{2 / 7}}
				{c^{10 / 7}\.v^{18 / 7}}
				\; .
				\label{eq:sigma-merge}
\end{align}
Setting $\sigma_{\rm s} \equiv \pi\.r_{\rm s}^{2}$, where $r_{\rm s}( M ) \equiv 2\.G M / c^{2}$ is the Schwarzschild radius of the mass $M$, we have
\begin{align}
	\sigma_{\rm mer}
		&\approx
				1.3\;\sigma_{\rm s}
				\left(
					\frac{m}{M}
				\right)^{\!2 / 7}
				\left(
					\frac{c}{v}
				\right)^{\!18 / 7}
				\; .
				\label{eq:sigma-merge-approx}
\end{align}
With $M \equiv M_{\rm SgrA} \approx 4 \times 10^{6}\.M_{\odot}$, $\sigma_{\rm s} \approx 5.1 \times 10^{14}\.{\rm km}^{2}$.
This leads to
\begin{align}
	\sigma_{\rm mer}
		&\approx
				1.1 \times 10^{27}\;
				\tilde{m}^{2 / 7}\.
				\tilde{v}^{- 18 / 7}\.
				{\rm km}^{2}
				\; ,
				\label{eq:sigma-merge-value-inserted}
\end{align}
where we have defined the dimensionless mass $\tilde{m} \equiv m / M_{\odot}$ and velocity $\tilde{v} \equiv v / ({\rm km / s})$.

The macro-SMBH merger rate is
\begin{align}
	\Gamma_{\rm mer} 
		&\equiv
				n\.\langle \sigma_{\rm mer}\.v \rangle_{v}^{}
				\; ,
				\label{eq:Gamma-merge}
\end{align}
where $n$ is the dark-matter number density at the Galactic center, and $\langle\,\cdot\,\rangle_{v}^{}$ denotes the average over an appropriate velocity distribution. 

The velocity distribution of dark matter near the Galactic center is very poorly known.
Purely for the sake of making a definite estimate, we take it to be Maxwellian:
\begin{align}
	g( v )\,\d^{3}v
		&\equiv
				\left(
					\frac{ 3 }{ 2 \pi\.v_{\rm RMS}^{2} }
				\right)^{\!3 / 2}
				\exp\!
				\left(
					-
					\frac{ 3\.v^{2} }{ 2\.v_{\rm RMS}^{2} }
				\right)
				\d^{3}v
				\, ,
				\label{eq:Maxwellian-velocity-distribution}
\end{align}
with a root-mean square velocity taken to be $v_{\rm RMS} \sim 100\.{\rm km / s}$ (with the scaling relative to this number retained throughout).

Unfortunately, the dark-matter energy density $\rho_{\rm DM}^{\rm center}$ at the Galactic center is also known quite poorly.
In our local neighborhood, $\rho_{\rm DM}^{\rm local} = 0.3\.{\rm GeV} / {\rm cm}^{3} \approx 10^{-42}\.M_{\odot} / {\rm km}^{3}$; 
however the value at the Galactic center is likely to be orders-of-magnitude larger (eg.~Ref.~\cite{Cirelli:2010xx}). 
We will therefore exhibit the functional dependence on the central dark-matter density explicitly, taking $10^{4}\.\rho_{\rm DM}^{\rm local}$ as our fiducial value.
For the Milky Way we find
\begin{widetext}
\begin{align}
	&\Gamma_{\rm mer}
		\approx
				1.8
				\left(
					\frac{ f_{\rm DM}\.\rho_{\rm DM} }{ 10^{4}\.\rho_{\rm DM}^{\rm local} }
				\right)\!
				\left(
					\frac{ m }{ 10^{-10}\.M_{\odot} }
				\right)^{\!-5 / 7}\!
				\left(
					\frac{ v_{\rm RMS} }{ 100\.{\rm km/s} }
				\right)^{\!-11 / 7}
				{\rm years}^{-1}
				\. ,
				\label{eq:Gamma-merge}
\end{align}
\end{widetext}
with $f_{\rm DM} \equiv \rho_{\rm macro} / \rho_{\rm DM}$ being the fraction of the dark-matter in macros.

Let us now turn to the question of detectability. 
We need to determine the expected amplitude for the gravitational waves. 
More precisely, we will focus on the root-mean square amplitude of the gravitational waves in their (dominant) second harmonic emitted toward infinity, at a time when the wave's frequency is $f_{2}$ \cite{Finn:2000sy},
\begin{align}
	h_{o,2}
		&\equiv
				\sqrt{
				\Big\langle
					{h_{2+}}^{\!2} + {h_{2 \times}\,}^{\!2}
				\Big\rangle \,}
				\label{eq:ho2}
				\; .
\end{align}
Above, $h_{2+}$ and $h_{2 \times}$ are the two waveforms, while $\langle\,\cdot\,\rangle$ denotes the average over a period of the waves, and over the directions of the waveforms (\cf~Ref.~\cite{Finn:2000sy} for details). 
Explicitly, Eq.~\eqref{eq:ho2} can be expressed as [\cf~Ref.~\cite{Finn:2000sy}, Eq.~(3.13)]
\begin{align}
	h_{o,2}
		&\approx
				\frac{ 1.7 \times 10^{-26} }{ r_{o} / 1{\rm Gpc} }\!
				\left( \frac{ m }{ M_{\odot} } \right)\!
				\left( \frac{ M }{ 100\.M_{\odot} } \right)^{\!\!2 / 3}\!\
				\left( \frac{ f_{2} }{ 10^{-3} {\rm Hz} } \right)^{\!\!2 / 3}
				,
				\label{eq:ho2-explicit}
\end{align}
which will be evaluate at the Earth, \ie~at a distance of $r_{o} \approx 8\.{\rm kpc}$ from the source.\footnote{
	In Eq.~\eqref{eq:ho2-explicit} a general-relativistic correction factor, 
	which depends on the distance of the inspiralling object from the central black hole, 
	as well as on the spin of the SMBH, has been included. 
	For the case of Sgr\.${\rm A}^{\!*}$, and a reasonable macro orbital radius, 
	its numerical value is approximately equal to $0.8$ 
	[see Ref.~\cite{Finn:2000sy}, Eq.~(3.12) and Tab.~III therein].}

The amplitude \eqref{eq:ho2-explicit} may be regarded as a 'bare' one, in the sense that it is not the one directly relevant for final signal evaluation. 
In fact, there are three types of enhancement of $h_{o,2}$: 
the first derives from observing many merger cycles of each inspiraling black hole; 
the second is due to the potential presence of many captured macros inspiralling simultaneously, and emitting gravitational-wave radiation at similar frequencies;
the third is due to the direction of the expected signal being known to be toward Sgr\.${\rm A}^{\!*}$.

During an observational time $\Delta t_{\rm obs}$, a gravitational-wave detector would experience $n_{\rm cycles} \approx f_{2}\.\Delta t_{\rm obs}$ cycles at frequency $f_2$.
So long as the time scale over which the gravitational-wave frequency changes is long compared to $\Delta t_{\rm obs}$, and assuming optimal signal processing, the so-called characteristic strain amplitude $h_{c, 2}$ is enhanced by approximately the square root of the number of cycles.

Meanwhile, depending on the merger rate, for a number $N$ of such merging black holes, one would get another enhancement factor of $\sqrt{N\,}$ 
(assuming incoherent superposition), leaving us with a characteristic strain amplitude
\begin{align}
	\tilde{h}_{c, 2}
		&\approx
				\sqrt{n_{\rm cycles}\,}\cdot
				\sqrt{N\,}\cdot
				h_{o,2}
				\; .
				\label{eq:hc2-tilde}
\end{align}
There is also an expected enhancement in $\tilde{h}_{c, 2}$ of order $\sqrt{4\pi /\Delta\Omega( f_{2} )\.}$, where $\Delta\Omega( f_{2} )$ is the angular resolution of the detector at the frequency $f_{2}$. 
However we will neglect this enhancement because it depends sensitively on the configuration of the detector.

Below, we will compare this characteristic strain amplitude $\tilde{h}_{c, 2}$ to recent sensitivity forecasts for LISA.
To do so, we must specify the frequency at which to compare to the proposed instrument's sensitivity. 
We take $f_{2}=(2 / 3)\.f_{2, {\rm isco}}$, where $f_{2, {\rm isco}}$ is the frequency of the wave's second harmonic at the innermost stable circular orbit \cite{Teukolsky:1973ha, Finn:2000sy}
\begin{align}
	r_{\rm isco}
		&=
				\frac{ r_{\rm s} }{ 2 }\!
				\left( 
					3 + Z_{2} - \frac{ a }{| a |}
					\sqrt{( 3 - Z_{1} )( 3 + Z_{1} + 2\.Z_{2})\,}
				\right)
				.
				\label{eq:rISCO}
\end{align}
Here
\begin{align}
	Z_{1}
		&\equiv
				1 + ( 1 - a^{2} )^{1/3}
				\left[ 
					( 1 + a )^{1/3} + ( 1 - a )^{1/3} 
				\right]
				\. ,
				\displaybreak[3]
				\\[1mm]
	Z_{2}
		&\equiv
				\big( 3\.a^{2} + {Z_{1}}^{2} \big)^{1/2}
				\, 
				\label{eq:risco}
\end{align}
and $a \equiv c\.J / G M^{2}$ is the central black hole's spin parameter. 
For ${\rm Sgr\.A^{\!*}}$ $a_{\rm SgrA} \approx 0.65$ \cite{2015PhyU...58..772D}, and 
\begin{align}
	f_{2, {\rm isco}}
		&=
				\frac{ 1 }{ \pi }
				\frac{ 2\.c }{ r_{\rm s} }\!
				\left[\mspace{-2mu}
					\left(
						\frac{ 2\.r_{\rm isco} }{ r_{\rm s} }
					\right)^{\!\! 3 / 2}
					\!\!\!\!+
					a\,
				\right]^{\! -1}
		\approx
				2 \times 10^{-3}\.{\rm Hz}
				\, .
				\label{eq:f2isco}
\end{align}
Hence, $f_{2} \simeq 1.3 \times 10^{-3}\.{\rm Hz}$.

The time the second-harmonic waves spends in the vicinity of frequency $f_{2}$ \cite{Finn:2000sy} is $\Delta t_{f} \approx 66 / \tilde{m}\,\.{\rm years}$. 
So, if we focus on macros with $m \leq M_{\odot}$, then $\Delta t_{f}$ is much larger than any feasible observational time.

We are now in a position to evaluate the doubly enhanced characteristic amplitude $\tilde{h}_{c, 2}$ suitable for comparison to the LISA sensitivity\footnote{
	The reason for focussing on Sgr\.${\rm A}^{\!*}$ can be found 
	by comparing the strain for Sgr\.${\rm A}^{\!*}$ at $8\.{\rm kpc}$ to that from, 
	or instance, a SMBH of $10^{9}\.M_{\odot}$ at a distance of one Mpc. 
	The latter amplitude is about a factor 20 smaller.}:
\begin{widetext}
\begin{align}
\begin{split}
	\tilde{h}_{c, 2}
		&\approx
				8.1 \times10^{-20}\.
				\left(
					\frac{ f_{\rm DM}\.\rho_{\rm DM} }{ 10^{4}\.\rho_{\rm DM}^{\rm local} }
				\right)^{\!\!1 / 2}\!
				\left(
					\frac{ m }{ 10^{-10}\.M_{\odot} }
				\right)^{\!\!1 / 7}\!
				\left(
					\frac{ v_{\rm RMS} }{ 100\.{\rm km/s} }
				\right)^{\!\!- 11/14}
				\left(
					\frac{ f_{2} }{ 10^{-3}\.{\rm Hz} }
				\right)^{\!\!- 4 / 3}
				\; .
				\label{eq:hc,2}
\end{split}
\end{align}
\end{widetext}
Comparing this to the proposed root-mean square noise level of LISA (as presented in Ref.~\cite{Finn:2000sy}), which is $h_{n} \approx 2 \times 10^{-21}$ (evaluated at $1.3 \times 10^{-3}\.{\rm Hz}$), we finds that LISA should detect gravitational-waves signals down to
\begin{align}
\begin{split}
	\tilde{m}_{\rm lower}
		&\approx
				8.3 \times 10^{-21}\.
				\left(
					\frac{ f_{\rm DM}\.\rho_{\rm DM} }{ 10^{4}\.\rho_{\rm DM}^{\rm local} }
				\right)^{\!\!-7 / 2}
				\left(
					\frac{ v_{\rm RMS} }{ 100\.{\rm km/s} }
				\right)^{\!\!11/2}
				,
				\label{eq:mu-limit}
\end{split}
\end{align}
which is given in units of solar masses.
Hence, for any sizable dark-matter fraction, $\tilde{m}_{\rm lower}$ is even smaller than the PBH evaporation threshold of $10^{15}\.{\rm g}$!

\begin{figure}
	\vs{-3mm}
	\centering
	\includegraphics[scale=1,angle=0]{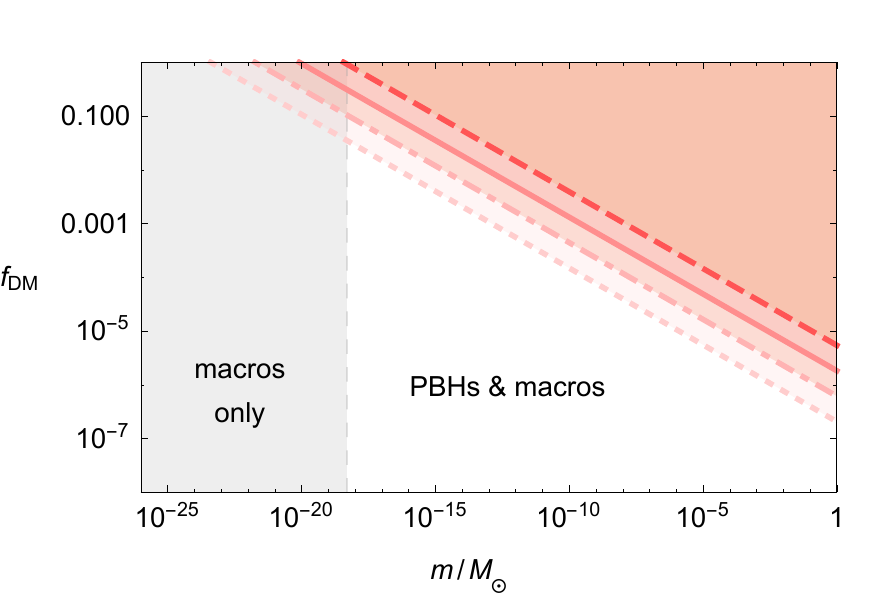}\;
	\vs{-6mm}
	\caption{
		Regions of potential observability of PBH/macroscopic dark-matter 
		gravitational-wave signals with LISA. 
		The observational period has been set to $10$ years and 
		we use a central dark-matter density of $10^{4}\.\rho_{\rm DM}^{\rm local}$.
		Depicted are lower boundaries of the dark-matter fraction 
		$f_{\rm DM} \equiv \rho_{\rm marco} / \rho_{\rm DM}$, as a function of PBH/macroscopic mass $m$ 
		in units of solar mass $M_{\odot}$; 
		a detection will be possible above these boundary values (red-shaded regions). 
		The different lines represent various root-mean square velocities
		assuming (for definiteness) a Maxwellian distribution of dark-matter velocities:
		$v_{\rm RMS} = 25\.{\rm km/s}$ (dotted line), 
		$v_{\rm RMS} = 50\.{\rm km/s}$ (dot-dashed line), 
		$v_{\rm RMS} = 100\.{\rm km/s}$ (solid line), 
		$v_{\rm RMS} = 200\.{\rm km/s}$ (dashed line). 
		}
	\label{fig:LISABound}
\end{figure}

Fig.~\ref{fig:LISABound} depicts the limits that LISA could put on macroscopic dark matter.
The existing open window for this form of dark matter ranging in density from black holes to just below atomic density{\,---\,}from $2 \times 10^{20}\.{\rm g}$ to $4 \times 10^{24}\.{\rm g}$, corresponding to $10^{-13}\.M_{\odot}$ to $2 \times 10^{-9}\.M_{\odot}${\,---\,}appears ripe for LISA.

A final concern is that tidal forces would rip apart the macro before it approached the SMBH horizon, at least partly invalidating our calculation of the gravitational-wave signal.
PBHs are certainly safe from this concern. 
Consider a less dense macro of size $d$. 
The tidal acceleration across a distance $d$ at the Schwarzschild radius of a Schwarzschild black hole of mass $M$ is \.{\rm g}$_{\rm tidal} \simeq ( d\.c^{2} / s_{\odot}^{2} )( M / M_{\odot})^{-2}$, where $s_{\odot}\simeq 3.0\.$km is the Schwarzschild radius of the Sun.
Compare this to the gravitational acceleration at the surface of a solar mass neutron star (NS), \.{\rm g}$_{\rm NS}\simeq 0.05\.c^{2} / s_{\odot}$.
Thus \.{\rm g}$_{\rm tidal} / g_{\rm NS} \simeq 22\.( d / s_{\odot} )( M / M_{\odot} )^{-2}$.
Nuclear-density macros with sizes less than the maximum allowed value within the open window, have $d / s_{\odot} \leq 3 \times 10^{-3}$, so clearly would not be disrupted.

For atomic-density macros, which would have $10^{6}\.{\rm cm} \sim d \sim 10^{9}\.{\rm cm}$, the situation is less optimistic, since \.{\rm g}$_{\rm tidal} \simeq 10^{10} ( d / {\rm cm})\.{\rm cm} / {\rm s}^{2}$.
Thus atomic-density macros would likely be ``spaghettified'' long before reaching the SMBH, significantly suppressing any gravitational-wave signal.
The precise dividing line between the robust nuclear-density macros and the disintegrating atomic-density ones would depend on the detailed physics that holds the macro together.

We therefore conclude that LISA has the potential to be a robust detector of primordial-black hole dark-matter candidate for their {\it entire} mass range! 
The same holds true for other macroscopic dark-matter candidates of approximately nuclear or higher density, but here even down to $\sim 10^{-20}\.M_{\odot}$.\\[-2mm]

%%%%%%%%%%%%%%%%%%%%%%%%%%%%%%%%%%%%%%%%%%%%%%%%%%%%%%%%%%%%

\acknowledgments
We indebted to Vitor Cardoso, Enrico Barausse, Emanuele Berti, and Paolo Pani for invaluable remarks on the evaluation of the characteristic gravitational-wave strain amplitude, correcting a mistake in the previous version of this manuscript.
We also thank Joe Bramante, Gil Holder, and Monica Valluri for helpful comments.
G.D.S thanks the Oskar Klein Center Cosmoparticle Physics for their hospitality
and FK thanks Case Western Reserve University and Lawrence Berkeley National Laboratory for their hospitality while this work was completed.
K.F.~and F.K.~acknowledge support from DoE grant DE-SC0007859 at the University of Michigan as well as support from the Michigan Center for Theoretical Physics. 
K.F.~and F.K.~acknowledge support by the Vetenskapsr{\aa}det (Swedish Research Council) through contract No.~638-2013-8993 and the Oskar Klein Centre for Cosmoparticle Physics.
G.D.S is partially supported by Case Western Reserve University grant DOE-SC0009946.

%%%%%%%%%%%%%%%%%%%%%%%%%%%%%%%%%%%%%%%%%%%%%%%%%%%%%%%%%%%%

\bibliography{refs}

\end{document}